\documentclass[11pt]{article}
\usepackage{cite}
\usepackage{amsmath,amsfonts,amssymb}
\usepackage[small,bf,hang]{caption}
\usepackage{graphics}
\usepackage{graphicx}
\DeclareGraphicsExtensions{.pdf}

\def\hybrid{
        \topmargin -20pt
        \oddsidemargin 0pt
        \headheight 0pt \headsep 0pt
        \textwidth 6.25in 
        \textheight 9.5in 
        \marginparwidth .875in
        \parskip 5pt plus 1pt \jot = 1.5ex}

\hybrid

 \csname
@addtoreset\endcsname{equation}{section}

\def\moth{\mathsurround=0pt}
\newdimen\zo \zo=0pt

\def\tick{\leaders\hrule height 0.5ex depth 0pt \hskip 0.5pt}
\def\upboxfill{$\moth \setbox\zo\hbox{\tick}%
  \hskip 3pt\hbox to 0pt{$\tick$\hss}\hrulefill \hbox to 7.5pt{$\tick$\hss}$}

\def\dtick{\leaders\hrule height .34pt depth 0.5ex \hskip 0.5pt}
\def\downboxfill{$\moth \setbox\zo\hbox{\dtick}%
  \hskip 2pt\hbox to 0pt{$\dtick$\hss}\hrulefill \hbox to 2pt{$\dtick$\hss}$}

\def\bec{\begin{center}}
\def\ec{\end{center}}

 \def\det{{\rm det\,}}
\def\be{\begin{equation}}
\def\ee{\end{equation}}
\def\bea{\begin{eqnarray}}
\def\eea{\end{eqnarray}}
\def\ba{\begin{array}}
\def\ea{\end{array}}

\begin{document}

\begin{titlepage}
\begin{center}
\vskip 2.5cm
{\Large \bf {
Holographic timelike entanglement and $c$ theorem for supersymmetric QFTs in ($ 0+1 $)d}}\\
\vskip 1.0cm
{\large {Dibakar Roychowdhury}}
\vskip 1cm
{\it {Department of Physics}}\\
{\it {Indian Institute of Technology Roorkee}}\\
{\it {Roorkee 247667, Uttarakhand, India}}\\

\vskip 2.5cm
{\bf Abstract}
\end{center}

\vskip 0.1cm

\noindent
\begin{narrower}
We present a holographic set up that computes timelike Entanglement Entropy (tEE) in $ (0+1) $d QFTs preserving some amount of SUSY. The first example we consider is that of $\mathcal{N}=2$ matrix models with massive deformations. These are dual to non-Abelian T-dual of $AdS_5 \times S^5$ that asymptotes to \emph{smeared} D0 branes. The second example, that we consider is of $ \mathcal{N}=4 $ superconformal quantum mechanical quivers in ($ 0+1 $)d that are dual to a class of type IIB backgrounds with an $ AdS_2 $ factor. In both of these examples, tEE reveals a remarkable similarity with holographic $ c $ function pertaining to a RG flow. We further compute the complexity in these models, which also reveals an identical behaviour indicating the fact that tEE is a measure of number of degrees of freedom for these ($ 0+1 $)d SQFTs in a RG flow from UV to deep IR.
\end{narrower}
\end{titlepage}
\newpage
\tableofcontents
\baselineskip=16pt
\section{Introduction and Overview}
Ryu-Takayanagi (RT) prescription for computing holographic entanglement entropy \cite{Ryu:2006bv}-\cite{Hubeny:2007xt} has gained renewed attention in the recent years, in the context of co-dimension one hyper-surface in the bulk, popularly known as the timelike entanglement entropy (tEE) \cite{Doi:2022iyj}-\cite{Afrasiar:2024ldn}. The purpose of the present paper is to show that tEE is indeed a good measure of the number of degrees of freedom in ($ 0+1 $)d SQFTs, while considering a RG flow from UV to deep IR. In other words, tEE can be associated with appropriate RG observables in ($ 0+1 $)d QFTs namely, the flow central charge \cite{Bea:2015fja}-\cite{Chatzis:2024kdu} as well as the complexity \cite{Susskind:2014rva}-\cite{Fatemiabhari:2024aua}. We show this taking two specific examples of SQFTs living in ($ 0+1 $)d, whose gravitational counterpart we elaborate below.  

The first example we consider is that of matrix models in ($ 0+1 $)d \cite{Lozano:2017ole}, that are dual to non-Abelian T dual (NATD) of $ AdS_5 \times S^5 $, where the T duality is performed along $ SU(2)\subset AdS_5 $. Global aspects of NATD \cite{delaOssa:1992vci} has been a challenge right from its beginning \cite{Alvarez:1993qi}-\cite{Alvarez:1994np}. The initial formulation of NATD were primarily confined to the NS sector, which were subsequently extended to the RR sector by authors in \cite{Sfetsos:2010uq}. This triggers a new realm for the AdS/CFT correspondence and ended up in producing a series of interesting papers \cite{Lozano:2011kb}-\cite{Roychowdhury:2023lxk}. 

One can produce two distinct classes of SQFTs through the application of NATD. In the first category, NATD is applied inside $ S^5 $, which results in a novel class of type IIA Giaotto-Maldacena (GM) backgrounds \cite{Gaiotto:2009gz} that are dual to $ \mathcal{N}=2 $ linear quivers \cite{Gaiotto:2009we} in ($ 1+3 $) dimensions. On the other hand, as mentioned above, when NATD is applied inside $ AdS_5 $, it produces a new class of type IIA background \cite{Lozano:2017ole} that are dual to irrelevant deformations of ($0+1$)d matrix models preserving $\mathcal{N}=2$ SUSY. These supersymmetric QFTs are non conformal in nature due to the presence of massive deformations, which results in a mass gap in the theory. RG flow in these matrix models corresponds to a flow central charge or $c$ function \cite{Macpherson:2014eza}, \cite{Bea:2015fja}-\cite{Chatzis:2024kdu} which decreases from UV to deep IR. One finds identical behaviour while computing the holographic complexity in these models \cite{Susskind:2014rva}-\cite{Fatemiabhari:2024aua}. As we show, both these entities reveal identical pole (of order $3$) stucture in the UV, as found in the case of the imaginary component of tEE.

The second example we consider is that of $ \mathcal{N}=4 $ superconformal quantum mechanics (SCQM) in ($ 0+1 $)d, that are dual to type IIB supergravity solutions containing an $ AdS_2 $ factor \cite{Lozano:2020txg}. These geometries are obtained through T duality inside $ AdS_3 $ of massive type IIA supergravity solutions\footnote{These massive type IIA solutions are obtained through NATD along internal directions of $ AdS_3 \times S^3 \times T^4 $.} \cite{Lozano:2019emq}. On the CFT side, the above operation corresponds to a dimensional reduction of $ \mathcal{N}=(0,4) $ SCFTs living ($ 1+1 $)d down to $ \mathcal{N}=4 $ SCQM in ($ 0+1 $)d. The above dimensional reduction preserves only one of the (right) sectors of the parent 2d SCFT, thereby producing only one central charge for the dimensionally reduced SCFT$ _1 $. 

As we show, tEE in $ \mathcal{N}=4 $ SCQM decreases along the RG flow from UV to IR, much similar in spirit to what has been found previously for matrix models. In particular, we establish precise relationship between holographic central charge ($c_{hol}$) and tEE near UV fixed point. Following the above line of arguments, the central charge must follow using a dimension reduction of $ \mathcal{N}=(0,4) $ SCFTs. Finally, we calculate complexity in $\mathcal{N}=4$ superconformal quivers and show that it could be mapped into tEE through holographic central charge near UV asymptotic.

The organisation for the rest of the paper is as follows. In section 2, we carry out a detailed calculation of tEE for $ \mathcal{N}=2 $ matrix models. We carry out similar analysis for $ \mathcal{N}=4 $ SCQM quivers in section 3. Finally, we conclude in section 4 along with some future remarks.
\section{$\mathcal{N}=2$ matrix models and tEE}
The purpose of this section is to discuss the 10d gravity set up that is dual to massive deformations of ($ 0+1 $)d matrix models preserving $ \mathcal{N} =2$ SUSY. These geometries are obtained via non-Abelian T duality (NATD) \cite{Lozano:2017ole} and could be recast as a solution in type IIA. The corresponding line element could be expressed in global coordinates as 
\begin{align}
\label{e2.1}
&ds_{10}^2 = L^2(-\cosh^2 r dt^2 +dr^2)+\frac{4 \alpha'^2 d\rho^2}{L^2 \sinh^2 r}\nonumber\\
&+\frac{4 \alpha'^2 L^2  \rho^2 \sinh^2r}{(16 \alpha'^2 \rho^2 +L^4 \sinh^4r)}(d\theta^2 + \sin^2\theta d\xi^2 )+L^2 ds^2_5
\end{align}
where $ ds^2_5 $ is the metric of a five sphere ($ S^5 $). Notice that, in the above line element \eqref{e2.1}, one encounters a spacetime singularity near $ r \sim 0 $, which is due to the presence of NS5 branes near the centre of NATD $ AdS_5 $. The metric \eqref{e2.1} is also accompanied by a background dilaton 
\begin{align}
e^{-2 \phi}=\frac{L^2 \sinh^2 r}{64 \alpha'^3}(16 \alpha'^2 \rho^2 +L^4 \sinh^4 r).
\end{align}

For our purpose, we introduce the following change of coordinate
\begin{align}
\cosh r = \frac{1}{\cos \gamma}
\end{align}
that results into the 10d line element and the dilaton of the form
\begin{align}
\label{e2.4}
&ds_{10}^2 = \frac{L^2}{\cos^2 \gamma}(-dt^2 +d\gamma^2)+\frac{4 \alpha'^2 \cos^2 \gamma}{L^2\sin^2\gamma}d\rho^2 \nonumber\\
&+ \frac{4 \alpha'^2 L^2 \rho^2 \sin^2 \gamma \cos^2 \gamma}{16 \alpha'^2 \rho^2 \cos^4 \gamma +L^4 \sin^4 \gamma}(d\theta^2 + \sin^2\theta d\xi^2 )+L^2 ds^2_5\\
&e^{- 2 \phi}=\frac{L^2 \tan^2 \gamma}{64 \alpha'^3}(16 \alpha'^2 \rho^2 +L^4 \tan^4 \gamma).
\label{e2.5}
\end{align}

Finally, we introduce the radial coordinate $ z= \cos \gamma  $ and reexpress the metric \eqref{e2.4} and the dilaton \eqref{e2.5} in the Poincare patch as
\begin{align}
\label{e2.6}
&ds_{10}^2 =\frac{L^2}{z^2}(-dt^2 + f(z)dz^2) +g(z) d\rho^2 +h(\rho ,z)(d\theta^2 + \sin^2\theta d\xi^2 )+L^2 ds^2_5\\
& e^{- 2 \phi}=\frac{L^2}{64 \alpha'^3}\frac{(1-z^2)}{z^2}(16 \alpha'^2 \rho^2 +\frac{L^4}{z^4} (1-z^2)^2).
\end{align}
where we define the above functions as
\begin{align}
\label{e2.8}
f(z)=\frac{1}{1-z^2}~;~g(z)= \frac{4 \alpha'^2 z^2}{L^2(1-z^2)}~;~h(\rho, z)=\frac{4 \alpha'^2 L^2 \rho^2 z^2 (1-z^2)}{16 \alpha'^2 \rho^2 z^4 +L^4 (1-z^2)^2}.
\end{align}

The radial coordinate $ z $ ranges between the deep IR $ z_{IR}=1 $ and the UV $ z_{UV}=0 $. On the other hand, the $ \rho $ direction is generated through the action of NATD on the original $ AdS_5 \times S^5 $, which is in principle unbounded. However, in a realistic set up, one needs to set an upper cut-off $\rho_{max}= \rho_0$, which in some true sense imposes a completion of the NATD background.

To proceed further, we take $ t=t(z) $, which results in the following induced metric on the co-dimension one surface that extends in the bulk
\begin{align}
ds_9^2 = \frac{L^2}{z^2}(-t'^2(z) + f(z))dz^2 +g(z) d\rho^2 +h(\rho ,z)(d\theta^2 + \sin^2\theta d\xi^2 )+L^2 ds^2_5.
\end{align}

The timelike entanglement entropy (tEE) is defined as \cite{Afrasiar:2024ldn}
\begin{align}
\mathcal{S}^{(tEE)}=\frac{1}{G_N}\int d^{9}x e^{- 2 \phi}\sqrt{\det g_9}.
\end{align}

A straightforward computation further reveals
\begin{align}
\label{e2.11}
\mathcal{S}^{(tEE)}=\frac{\pi V_5 L^4 \rho^3_0}{ 24 G_N} \int_0^1 dz \chi(z)\sqrt{f(z) -t'^2(z)}~;~\chi(z)=\frac{1}{z^4}(1-z^2)^{3/2}
\end{align}
where $ V_5 $ is the volume of the internal five sphere.

The equation of motion, that readily follows from \eqref{e2.11} can be expressed as
\begin{align}
\label{e2.12}
t'^2(z) = \frac{C^2 f(z)}{C^2 + \chi^2 (z)}
\end{align}
where $ C $ is the constant of integration.

Substituting \eqref{e2.12} into \eqref{e2.11}, the tEE finally reads as
\begin{align}
\label{e2.13}
\mathcal{S}^{(tEE)}=\frac{\pi V_5 L^4 \rho^3_0}{ 24 G_N} \int_0^1 dz\frac{\chi^2 (z)\sqrt{f(z)}}{\sqrt{C^2 + \chi^2(z)}}.
\end{align}

Given \eqref{e2.13}, we do have the following three conditions at our disposal. The first two classes correspond to a pair of connected surfaces with and without a turning point ($ z=z_0 $) in the bulk. These surfaces yield UV divergences that can be tamed by introducing a pair of disconnected surfaces which refers to the third category \cite{Afrasiar:2024ldn}. Below, we discuss each of these cases in detail.
\subsection{Connected surfaces with turning point}
\begin{figure}
\begin{center}
\includegraphics[scale=0.9]{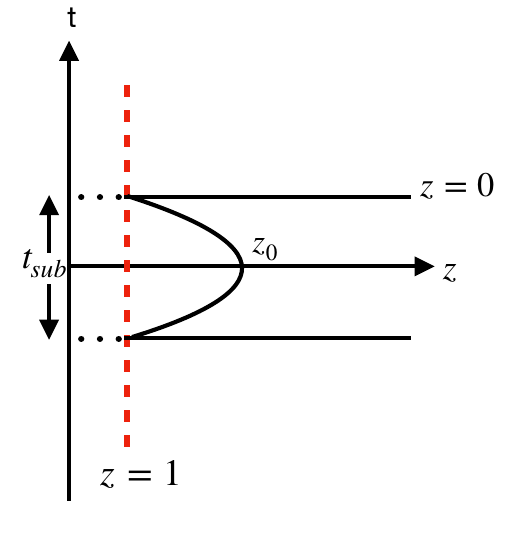}
  \caption{Extremal connected surface of type I with turning point at $ z=z_0 $.} \label{fig1}
  \end{center}
\end{figure}

This corresponds to a pair of connected surfaces that are glued together at the turning point $ z=z_0 $ (see Fig.\ref{fig1}), where $ 0<z_0 < 1 $. The corresponding boundary conditions read as
\begin{align}
t'(z=z_0)=\pm \infty ~;~ t'(z=1)=\pm \infty.
\end{align}

Clearly, the condition at $ z=1 $ is trivially satisfied due to the presence of the function $ f(z) $ in \eqref{e2.12}, which diverges in the deep IR ($ z \sim 1 $). On the other hand, in order to figure out the turning point at $z=z_0 $, we redefine $ C=i \tilde{C} $, such that $ C^2<0 $. This yields $ \tilde{C}^2 = \chi^2_0 $, where $ \chi_0=\chi(z)|_{z=z_0} $. This leads to the following tEE pertinent to the connected surface of type I
\begin{align}
\label{e2.15}
\mathcal{S}^{(tEE)}_{con}\equiv \mathcal{S}^{(Im)}_{con}=\frac{-i \pi V_5 L^4 \rho^3_0}{ 12 G_N} \int_{z_0}^1 dz \frac{\chi^2 (z)\sqrt{f(z)}}{\sqrt{ |\chi_0^2-\chi^2(z)|}}.
\end{align}

Length of the corresponding subsystem is given by
\begin{align}
\label{e2.16}
t_{sub} =2 \int_{z_0}^1dz\frac{|\chi_0|\sqrt{f(z)}}{\sqrt{ |\chi_0^2-\chi^2(z)|}}.
\end{align}

An exact analytic evaluation of \eqref{e2.15} is a difficult task. However, at the first place, it is possible to figure out \emph{asymptotic} solutions pertaining to \eqref{e2.15}. For example, one could think of expanding \eqref{e2.15} about $ z_0 \simeq z_{IR}=1$, which yields at LO
\begin{align}
\label{e2.17}
\mathcal{S}^{(Im)}_{con}|_{z_0 \sim 1}=\frac{-i \pi V_5 L^4 \rho^3_0}{ 36 G_N} \left(2+\frac{1}{z_0^3}-\frac{3}{z_0}\right)\Big|_{z_0 \sim 1}=0.
\end{align}

\begin{figure}
\begin{center}
\includegraphics[scale=0.5]{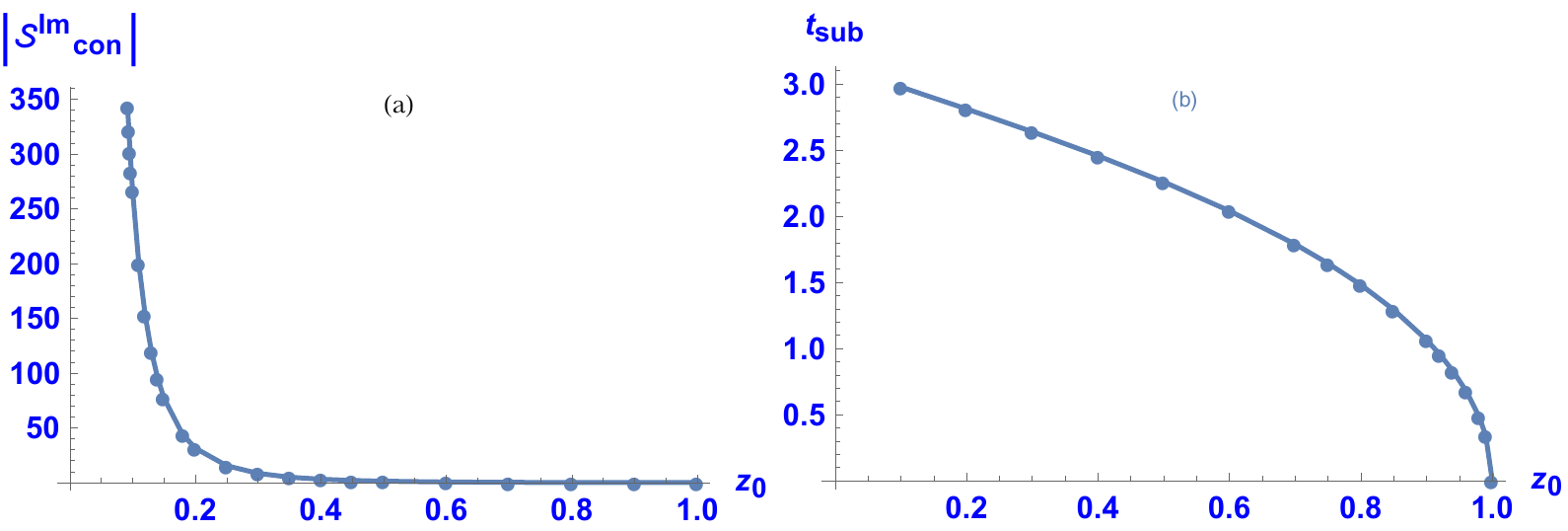}
  \caption{(a) Plot of imaginary tEE \eqref{e2.15} with the location of the turning point ($ z_0 $). (b) Plot of subsystem size \eqref{e2.16} with the location of the turning point ($ z_0 $).} \label{figst1}
  \end{center}
\end{figure}

As a second approximation, one might consider an expansion near the UV asymptotic ($ z_{UV}= 0 $) of the spacetime, where we take the turning point close to the boundary namely $ z_0 \sim \epsilon \sim 0$, where $ \epsilon $ is some appropriate UV cut-off. This yields the following tEE 
\begin{align}
\label{e2.18}
\mathcal{S}^{(Im)}_{con}|_{z_0 \sim \epsilon}=\frac{-i \pi V_5 L^4 \rho^3_0}{ 84 G_N \epsilon^3_0} (1-\epsilon^2_0)^{7/2} = -i\mathcal{S}_0 \Big(\frac{1}{\epsilon^3} -\frac{7}{2 \epsilon}+\frac{35 \epsilon}{8}-\frac{35 \epsilon^3}{16}+\cdots\Big)
\end{align}
where $ \mathcal{S}_0 =\frac{ \pi V_5 L^4 \rho^3_0}{ 84 G_N } $ is the overall pre-multiplicative constant. Clearly, one could see that the imaginary component diverges as we move towards UV asymptotic of the spacetime. This could be verified as an \emph{exact} function of the turning point ($ 0<z_0<1 $) using numerics (see Fig.\ref{figst1}(a)). 

As Fig.\ref{figst1}(a) shows, the imaginary contribution vanishes in the deep IR ($ z_0 \sim 1 $) and diverges near the UV asymptotic ($ z_0 \sim 0 $). These are precisely reflected in \eqref{e2.17} and \eqref{e2.18}. The divergence above is understandable, as the corresponding area functional pertaining to type I extremal surface is not \emph{renormalised}. In other words, pushing the turning point ($ z_0 $) towards asymptotic infinity results in an infinite area contribution and hence a divergent tEE. The renormalisation can be done by adding appropriate counter term, whose role is typically played by the pair of disconnected (type III) surfaces, to be elaborated in the subsequent section.

On a similar note, one could estimate the subsystem length \eqref{e2.16} in both asymptotic limits. Considering a deep IR approximation, one finds
\begin{align}
\label{e2.19}
t_{sub}|_{z_0 \sim 1} = 2 |\chi_0|\int_{z_0}^{1}dz\frac{z^4}{(1-z^2)^2}=0
\end{align}
which is subjected to the fact that the integral goes like $ \mathcal{O}(\sqrt{1-z^2_0} )$, in the limit $ z_0 \sim 1 $. 

On the other hand, in the near boundary limit $ z_0 \sim 0 $, one finds
\begin{align}
\label{e2.20}
t_{sub}|_{z_0 \sim 0} = 2 \int_{z_0}^{1}\frac{dz}{\sqrt{1-z^2}}=\pi
\end{align}
which saturates to a finite value as the tip ($ z_0 $) of the extremal surface reaches the boundary. 

\begin{figure}
\begin{center}
\includegraphics[scale=0.45]{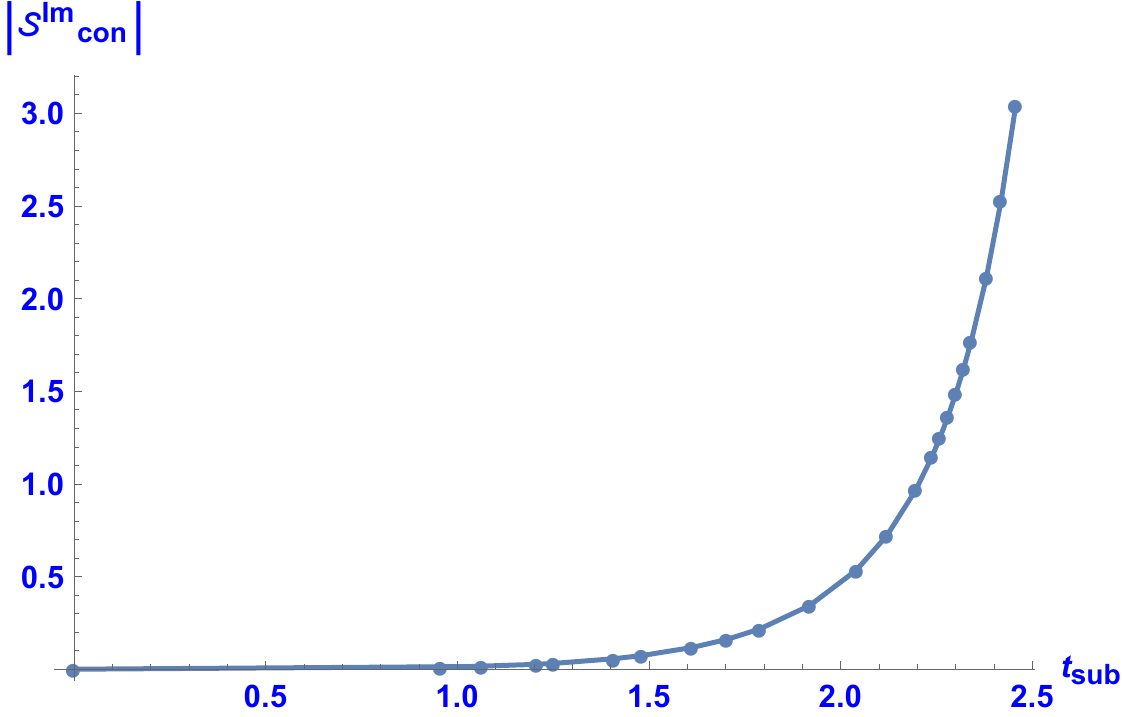}
  \caption{Plot of imaginary component \eqref{e2.15} of tEE with the size \eqref{e2.16} of the entangling region.} \label{figst2}
  \end{center}
\end{figure}

In Fig.\ref{figst1}(b), we estimate the subsystem length \eqref{e2.16} as an exact function of the turning point ($ 0<z_0 <1$), which clearly confirms our analytical findings in \eqref{e2.19} and \eqref{e2.20}. Combining the above pictures together, one could express $ |\mathcal{S}^{(Im)}_{con}| $ as a function of the subsystem size \eqref{e2.16}. This is depicted in Fig.\ref{figst2}. As one could see, tEE \eqref{e2.15} remains almost uniform when the system size is smaller than a critical dimension. Beyond this critical value, the entanglement \eqref{e2.15} increases at a faster rate and thereby reaching infinity when the system size \eqref{e2.16} saturates.
\subsection{Connected surfaces without a turning point}
\begin{figure}
\begin{center}
\includegraphics[scale=0.9]{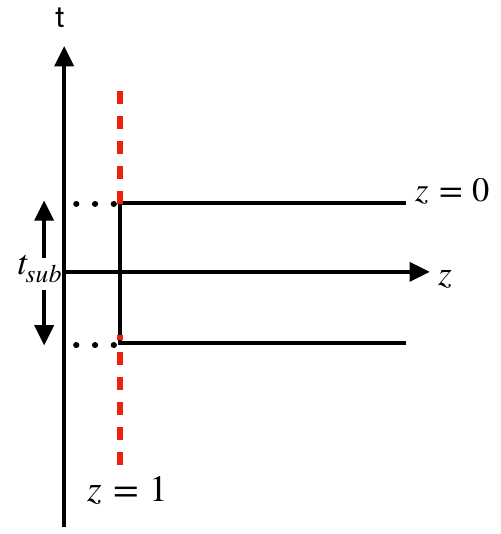}
  \caption{A pair of connected surfaces without a turning point.} \label{fig2}
  \end{center}
\end{figure}

These are type II extremal surfaces that correspond setting $ C^2=\chi^2_0>0 $, which does not possess a real root for the denominator in \eqref{e2.12} and hence there are no turning points in the bulk (see Fig.\ref{fig2}). The integral is UV divergent for which we need a UV regulator. 

The associated boundary conditions read as
\begin{align}
t'(z=0)=0 ~;~ t'(z=1)=\pm \infty.
\end{align}

As one can see, these boundary conditions in \eqref{e2.17} are trivially satisfied by \eqref{e2.12}. For example, in the UV $ z \sim 0 $, the functions above behave as
\begin{align}
f(z \sim 0)=1~;~ \chi (z \sim 0)\sim \frac{1}{z^4}
\end{align}
which thereby satisfies the boundary condition in the UV namely, $ t'(z \sim 0)\sim z^4 \sim 0 $.

The tEE associated with type II extremal surfaces is given by
\begin{align}
\label{e2.23}
\mathcal{S}^{(tEE)}_{con}\equiv \mathcal{S}^{(Re)}_{con}=\frac{\pi V_5 L^4 \rho^3_0}{ 12 G_N} \int_0^1 dz\frac{\chi^2 (z)\sqrt{f(z)}}{\sqrt{C^2 + \chi^2(z)}}.
\end{align}

This is accompanied by a subsystem whose length is given by
\begin{align}
t_{sub}= t^{(II)}=2 \int_0^1 dz \frac{C\sqrt{f(z)}}{\sqrt{\chi^2(z)+C^2}}.
\end{align}

One can simplify \eqref{e2.23} further by substituting the function $ \chi(z) $, which yields
\begin{align}
\mathcal{S}^{(Re)}_{con}=\frac{\pi V_5 L^4 \rho^3_0}{ 12 G_N} \int_\epsilon^1 \frac{dz}{z^4}\frac{(1-z^2)^{5/2}}{\sqrt{C^2 z^8 +(1-z^2)^3}}
\end{align}
where $ \epsilon $ is the UV cut-off. Considering $ C^2 \ll 1 $, one finds an approximate function
\begin{align}
\label{e2.26}
\mathcal{S}^{(Re)}_{con}=\frac{\pi V_5 L^4 \rho^3_0}{ 12 G_N}  \int_\epsilon^1 \frac{dz}{z^4}(1-z^2)+\mathcal{O}(C^2 z^4)
\end{align}
which separates the UV divergence from the finite part. Clearly, the leading term contributes a divergent piece in the asymptotic limit and combining both we find
\begin{align}
\label{e2.27}
\mathcal{S}^{(Re)}_{con}=\frac{\pi V_5 L^4 \rho^3_0}{ 36 G_N} \Big(2+\frac{1}{\epsilon^3} \Big).
\end{align}

\begin{figure}
\begin{center}
\includegraphics[scale=0.9]{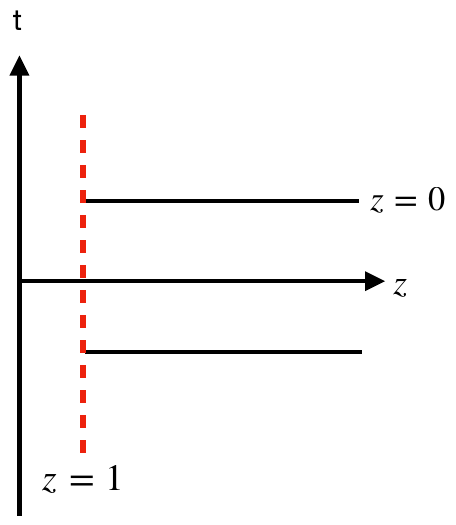}
  \caption{A pair of disconnected surfaces that acts like a UV regulator.} \label{fig3}
  \end{center}
\end{figure}

From \eqref{e2.18} and \eqref{e2.27}, it is evident that both imaginary and real components of tEE suffer from a UV divergence, which exhibits a pole of order $ 3 $ at asymptotic infinity. These divergences can be tamed\footnote{As we argue, the divergence in \eqref{e2.15} actually counts the number of degrees of freedom in a RG flow as this has a feature unique to that of a $c$ function. On the other hand, the real component \eqref{e2.27} always diverges and can be exactly cancelled by adding counter term \eqref{e2.29}, leaving only the imaginary component \eqref{e2.15}.}, typically by introducing an appropriate counter term \cite{Afrasiar:2024lsi}. These are type III surfaces that comprise of a pair of disconnected surfaces as shown in Fig.\ref{fig3}. This corresponds to setting $ C^2=0 $ in \eqref{e2.13}, which yields the following boundary conditions
\begin{align}
t'(z=0)=0 ~;~ t'(z=1)=0.
\end{align}

The associated extremal surface yields the tEE of the following form
\begin{align}
\label{e2.29}
\mathcal{S}^{(III)}_{discon}=\frac{\pi V_5 L^4 \rho^3_0}{ 12 G_N} \int_0^1 dz \chi(z)\sqrt{f(z)}
\end{align}
which is precisely the leading term as found in \eqref{e2.26}. Therefore, by adding this counter term with some appropriate pre factor, one should get rid of the UV divergences in tEE. 
\subsection{Comments on $c$ function}
As we have noticed before, the imaginary component \eqref{e2.15} of tEE diverges in the UV and vanishes in the deep IR. In what follows, in this section, we show that the above behaviour can be associated with a flow central charge \cite{Macpherson:2014eza}, \cite{Bea:2015fja}-\cite{Chatzis:2024kdu} in a holographic RG flow. This turns out to be a nice field theoretic aspect of tEE \eqref{e2.15}, where one can associate time entanglement with the number of degrees of freedom associated with the matrix model in a RG flow and this might well be the generic feature for other QFTs in diverse dimensions.

For generic QFTs, the $c$ function captures the number of degrees of freedom in a RG flow, which is a \emph{monotonically} decreasing function (of energy) as one moves from UV to deep IR. In case the RG flow hits fixed points, the entity takes constant values both for the UV and IR CFTs, such that $c_{UV}>c_{IR}$. For the purpose of our paper, we follow the proposal for constructing a flow central charge (or $c$ function) based on the earlier works of \cite{Macpherson:2014eza}, \cite{Bea:2015fja}-\cite{Chatzis:2024kdu}. 

We rewrite the background metric \eqref{e2.1} in the global coordinates as
\begin{align}
&ds_{10}^2=L^2(-\cosh^2 r dt^2 +dr^2)+\mathcal{G}_{ij}d\omega^{i}d\omega^{j}\nonumber\\
&\mathcal{G}_{ij}d\omega^{i}d\omega^{j}=\frac{4 \alpha'^2 d\rho^2}{L^2 \sinh^2 r}+\frac{4 \alpha'^2 L^2  \rho^2 \sinh^2r}{(16 \alpha'^2 \rho^2 +L^4 \sinh^4r)}(d\theta^2 + \sin^2\theta d\xi^2 )+L^2 ds^2_5
\end{align}
where, $ i,j(=1,\cdots ,8) $ denote the indices of the (eight dimensional) internal manifold.

Following \cite{Chatzis:2024kdu}, the flow central charge for a ($ d+1 $) QFT can be expressed as
\begin{align}
\label{e2.31}
c_{flow}=\frac{d^d \beta^{d/2}\mathcal{H}^{(2d+1)/2}}{G_N (\mathcal{H'})^d}~;~\sqrt{\mathcal{H}}=\int d^n\omega e^{-2\phi}\sqrt{\det \mathcal{G}_{ij}}
\end{align}
where the integration runs over the internal directions of the manifold and the function $\beta(r)$ is associated with the metric coefficient ($g_{rr}$) along the radial coordinate ($r$).

The NATD background, that asymptotes to smeared D0 branes, is conjectured to be dual to a  ($ 0+1 $)d matrix model with massive deformations \cite{Lozano:2017ole}. In other words, considering $ d=0 $, one can introduce an analogue flow function for the dual matrix model 
\begin{align}
c_{flow}= \frac{1}{G_N}\int d^8 \omega e^{-2\phi}\sqrt{\det \mathcal{G}_{ij}}.
\end{align}

A straightforward computation further reveals
\begin{align}
\label{e2.33}
c_{flow}= \frac{\pi \rho^3_0 L^3V_5}{6G_N}\sinh^3r =\frac{\pi \rho^3_0 L^3V_5}{6G_N}\frac{(1-z^2)^{3/2}}{z^3}
\end{align}
where in the last step we use the map that relates global coordinates to Poincare patch.

\begin{figure}
\begin{center}
\includegraphics[scale=0.45]{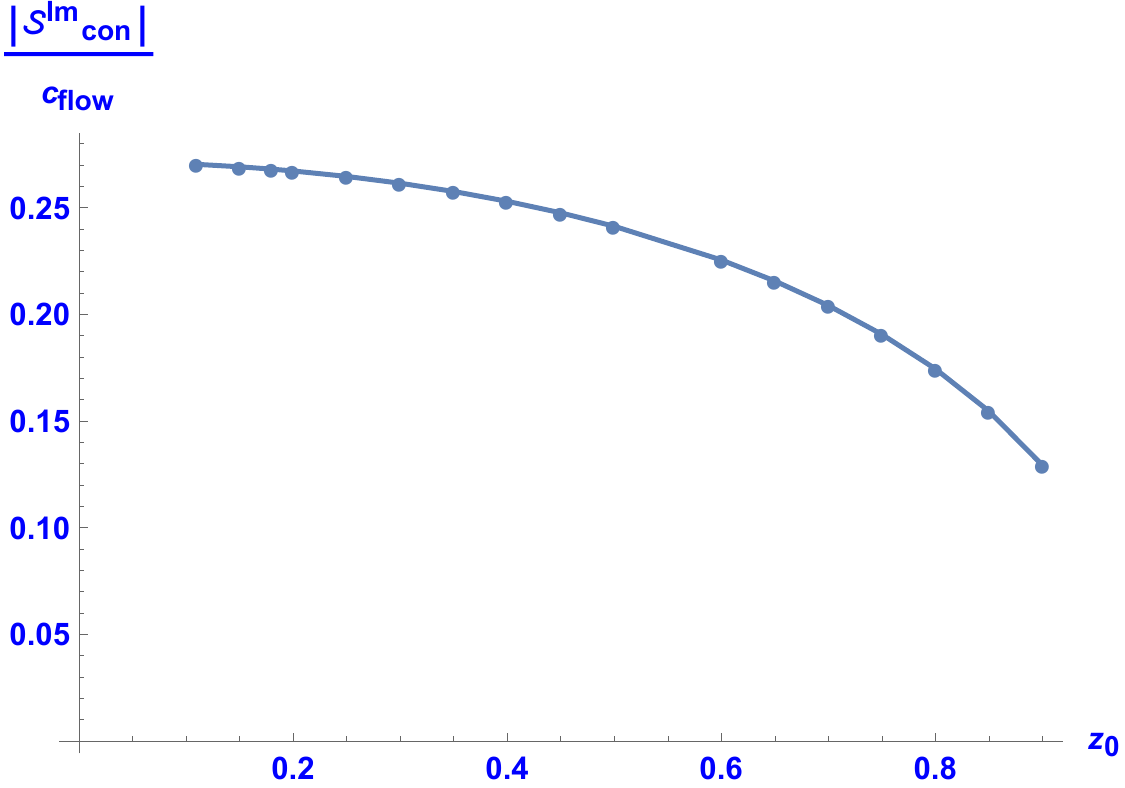}
  \caption{In the above Fig.\ref{ttevscflow}, we compare tEE \eqref{e2.15} with the flow central charge \eqref{e2.33} by taking their ratio as a function of the RG coordinate $ z=z_0 $. As we move from the deep interior ($z_0 \sim 1$) to the boundary ($z_0 \sim 0$), the ratio increases and eventually saturates to a bound \eqref{e2.34}. From the plot, one could further infer that at the initial stage ($ z_0 \sim 1 $) of the RG flow, the flow central charge increases at a rate faster than the tEE. On the other hand, at a later stage ($ z_0 \sim 0 $), the rate of growth of both entities becomes equal and the ratio saturates.} \label{ttevscflow}
  \end{center}
\end{figure}

It is interesting to notice that \eqref{e2.33} precisely reproduces the behaviour of \eqref{e2.15} (see Fig.\ref{figst1}(a)), namely it vanishes in the deep IR ($ z_{IR} =1 $) and diverges in the UV ($ z_{UV}=0 $). In particular, it exhibits a pole (of order $ 3 $) identical to that of \eqref{e2.18}, which leads us towards the following relationship between the $ c $ function and the tEE, near the UV asymptotic ($ z \sim  0 $)
\begin{align}
\label{e2.34}
|\mathcal{S}^{(Im)}_{con}|=\frac{L}{14}c_{flow}.
\end{align}
The above relation clearly reflects that tEE could be thought of as measure of number of degrees of freedom much like that of a flow central charge. In other words, the entity \eqref{e2.15} indeed can be taken as a measure of the degrees of freedom for ($ 0+1 $)d QFTs. 
\subsection{A note on complexity}
As the authors in \cite{Chatzis:2024kdu}, \cite{Fatemiabhari:2024aua} argue that complexity \cite{Susskind:2014rva}-\cite{Stanford:2014jda} in QFTs could be a good measure of number of degrees of freedom in the system, therefore given the above scenario, it seems to be a well defined platform to check this statement in the context of matrix models in ($ 0+1 $)d. The quantum computational complexity is a measure of minimum number of elementary quantum gates required in a quantum circuit in order to build up a generic state ($ \vert \Psi \rangle$) in the Hilbert space ($ \mathcal{H} $), starting from a reference state. In what follows, we would consider the CV conjecture as posited in \cite{Stanford:2014jda}, that refines the original idea of \cite{Susskind:2014rva}, by stating that the complexity ($ \mathcal{C} $) in the dual QFT at a particular time can be computed by knowing the volume of a maximal spacelike slice in the bulk, ending on a boundary. 

The general idea is to express the 10d space time \eqref{e2.1} as 
\begin{align}
ds^2_{10}= g_{tt}dt^2 +ds^2_9
\end{align}
where the nine dimensional metric can be expressed as
\begin{align}
ds^2_9 = L^2dr^2+\frac{4 \alpha'^2 d\rho^2}{L^2 \sinh^2 r}+\frac{4 \alpha'^2 L^2  \rho^2 \sinh^2r}{(16 \alpha'^2 \rho^2 +L^4 \sinh^4r)}(d\theta^2 + \sin^2\theta d\xi^2 )+L^2 ds^2_5.
\end{align}

The complexity ($ \mathcal{C} $) associated with the state $ \vert \Psi \rangle \in  \mathcal{H} $ in the dual CFTs could be computed using the holographic prescription \cite{Chatzis:2024kdu}, \cite{Fatemiabhari:2024aua}, which states that it is the maximal volume associated with the spacelike hyper-surface at a fixed time $ t=t_0 $ and is given by 
\begin{align}
\label{e2.37}
\mathcal{C}_V = \frac{1}{G_N}\int d^9x \frac{e^{-2 \phi}}{\sqrt{\mathcal{A}}}\sqrt{\det g_9}
\end{align}
where the conformal factor $ \mathcal{A}=1 $ for the present example.

A straightforward computation further reveals
\begin{align}
\label{e2.38}
\mathcal{C}_V &= \frac{\pi \rho_0^3 L^4 V_5}{6 G_N}\int_{0}^{\infty}dr \sinh^3 r \nonumber\\
&= \frac{\pi \rho_0^3 L^4 V_5}{6 G_N}\int_{\epsilon}^{1} dz \frac{(1-z^2)}{z^4}\nonumber\\
&=\frac{\pi \rho_0^3 L^4 V_5}{18G_N}\Big( 2+\frac{1}{\epsilon^3}\Big)
\end{align}
where $\epsilon$ is the UV cut-off, as mentioned previously.

Clearly, near the UV asymptotic ($z\sim \epsilon$), complexity \eqref{e2.38} reveals a pole structure identical to that of tEE \eqref{e2.18} and these two entities could be related to each other
\begin{align}
|\mathcal{S}^{(Im)}_{con}|_{UV}=\frac{3}{14}(\mathcal{C}_V-\mathcal{C}_V^{(0)})
\end{align}
where $\mathcal{C}_V^{(0)} = \frac{\pi \rho_0^3 L^4 V_5}{9G_N}$ is the rescaled complexity, that vanishes in the IR ($z=1$) and exhibits a pole of order $3$ near UV asymptotic ($z\sim \epsilon$).
\subsection{tEE in Euclidean signature}
So far our analysis was restricted to Lorentzian framework. We extend it to Euclidean signature by rescaling $ t = i t_E $, which leads to the following metric
\begin{align}
ds_9^2 = \frac{L^2}{z^2}(t_E'^2(z) + f(z))dz^2 +g(z) d\rho^2 +h(\rho ,z)(d\theta^2 + \sin^2\theta d\xi^2 )+L^2 ds^2_5
\end{align}
and the associated tEE in the Euclidean framework
\begin{align}
\label{e2.38}
\mathcal{S}^{(tEE)}_E=\frac{\pi V_5 L^4 \rho^3_0}{ 12 G_N} \int_{z_0}^0 dz \chi(z)\sqrt{f(z) +t_E'^2(z)}.
\end{align}

The equation of motion that follows from \eqref{e2.38}, could be formally expressed as
\begin{align}
\label{e2.39}
|t'_E(z)|=\frac{\chi_0 \sqrt{f (z)}}{\sqrt{\chi^2 (z)- \chi^2_0}}
\end{align}
where $ \chi_0 $ corresponds to a turning point at $ z=z_0 $. 

The corresponding subsystem length is given by
\begin{align}
\label{e2.43}
t_E=2\int_{z_0}^0 dz \frac{\chi_0 \sqrt{f (z)}}{\sqrt{\chi^2 (z)- \chi^2_0}}.
\end{align}

As the turning point approaches deep IR ($ z_0 \sim 1 $), one could approximate \eqref{e2.43} to find
\begin{align}
\label{e2.44}
t_E=2 \chi_0 \int_{z_0 \sim 1}^0 dz \frac{ \sqrt{f (z)}}{\chi(z)}=2 \chi_0 \int_{z_0 \sim 1}^0 dz \frac{z^4}{(1-z^2)^2}=0.
\end{align}

Using \eqref{e2.39}, the tEE \eqref{e2.38} finally reads as
\begin{align}
\label{e2.45}
\mathcal{S}^{(tEE)}_E=\frac{\pi V_5 L^4 \rho^3_0}{ 12 G_N}\int_{z_0}^0 dz \frac{\chi^2(z) \sqrt{f (z)}}{\sqrt{\chi^2 (z)- \chi^2_0}}.
\end{align}

Typically, the above integral \eqref{e2.45} diverges in both deep IR ($ z_0 \sim 1 $) as well as asymptotic infinity ($ z_0 \sim 0 $). Therefore, one needs to add a counter term in order to obtain a finite answer. This yields the following regularised tEE in Euclidean signature
\begin{align}
\label{e2.46}
\mathcal{S}_{reg}=\mathcal{S}^{(tEE)}_E -\frac{\pi V_5 L^4 \rho^3_0}{ 12 G_N} \int_{z_0}^0 dz \chi(z)\sqrt{f(z)}.
\end{align}

\begin{figure}
\begin{center}
\includegraphics[scale=0.4]{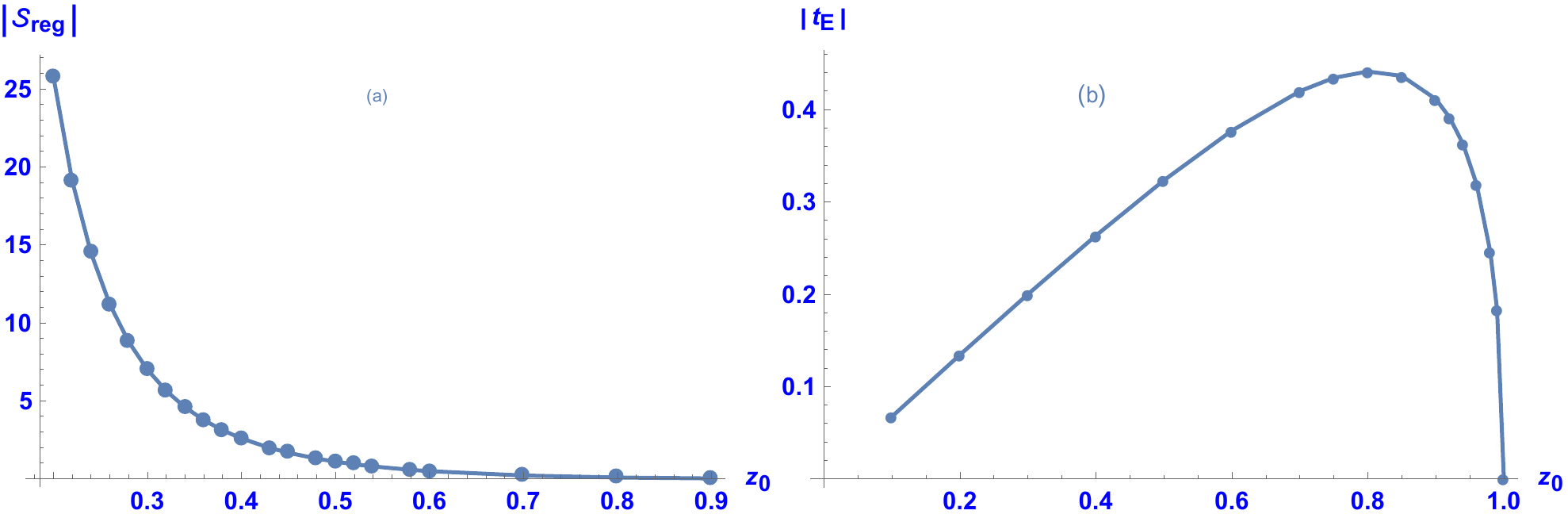}
  \caption{(a) Plot of Euclidean tEE \eqref{e2.46} with the location of the turning point ($ z_0 $). (b) Plot of subsystem size \eqref{e2.43} with the location of the turning point ($ z_0 $).} \label{figest}
  \end{center}
\end{figure}

As it turns out, (regularised) tEE \eqref{e2.46} increases with the as the turning point ($ z_0 $) approaches the boundary (see Fig.\ref{figest}(a)). This is an indicative of the fact that the area of the (regularised) connected extremal surface increases near the UV asymptotic \cite{Afrasiar:2024lsi}, which is identical in spirit to that of tEE \eqref{e2.15} in Lorentzian signature (Fig.\ref{figst1}(a)). The subsystem size \eqref{e2.43}, on the other hand, reaches a maxima for some intermediate $ z_0=z_{max} $ (where $0< z_{max}<1 $) and thereby decreases for $ z_0 < z_{max} $, reaching a zero near asymptotic infinity (see Fig.\ref{figest}(b)) \cite{Afrasiar:2024lsi}.
\section{$\mathcal{N}=4$ conformal quantum mechanics and tEE}
Next, we move on to type IIB backgrounds preserving an $ AdS_2 $ factor, that are dual to a class of $\mathcal{N}=4$ conformal quantum mechanics (SCQM) in ($ 0+1 $)d \cite{Lozano:2020txg}. The seed background for these class of geometries are considered to be massive IIA supergravity solutions with an $ AdS_3 $ factor, that are dual to $ \mathcal{N}=(0,4) $ SCFTs in ($ 1+1 $)d \cite{Lozano:2019emq}. The $ AdS_2 $ supergravity is obtained following a T-duality inside $ AdS_3 $, which results in the following 10d metric
\begin{align}
\label{e3.1}
&ds^2_{10}=f_1(\rho)(- \cosh^2r dt^2 +dr^2)+ds^2_8\\
&ds^2_{8}=f_2(\rho)(d \theta^2 +\sin^2\theta d\xi^2)+f_3(\rho)ds^2_{CY_2}+f_4(\rho)(d\rho^2 + d\psi^2)
\end{align}
where the individual metric components may be identified as
\begin{align}
&f_1(\rho)=\frac{u(\rho)}{4\sqrt{\hat{h}_4(\rho) h_8(\rho)}}~;~f_2(\rho)=4 f_1(\rho)\frac{\hat{h}_4 (\rho) h_8(\rho)}{4\hat{h}_4 (\rho) h_8(\rho)+(u'(\rho))^2}\\
&f_3(\rho)=\sqrt{\frac{\hat{h}_4(\rho)}{h_8(\rho)}}~;~f_4(\rho)=\frac{\sqrt{\hat{h}_4(\rho)h_8(\rho)}}{u(\rho)}.
\end{align}

As special choice, we consider the following (NATD) solution, which sets
\begin{align}
u (\rho)=\rho~;~\hat{h}_4 (\rho)= \rho~;~h_8 (\rho)=\rho
\end{align}
and thereby satisfying the BPS equations as well as preserving the background SUSY \cite{Lozano:2020txg}
\begin{align}
u'' (\rho)=0~;~\hat{h}_4'' (\rho)= 0~;~h_8'' (\rho)=0.
\end{align}

With these choices, the metric coefficients simplify and turn out to be
\begin{align}
f_1(\rho)=\frac{1}{4}~;~f_2(\rho)=\frac{\rho^2}{4\rho^2+1}~;~f_3(\rho)=1~;~f_4(\rho)=1
\end{align}
which will serve the purpose for our subsequent analysis.

Finally, the dilaton for the above NATD background \eqref{e3.1} reads as
\begin{align}
e^{-2 \phi}=\frac{1}{4}(4 \rho^2+1).
\end{align}
\subsection{tEE in Lorentzian signature}
Considering $ t=t(r) $, the metric of the codimension one surface reads as
\begin{align}
ds^2_9=\frac{1}{4}(1-\cosh^2 r t'^2(r))dr^2 + ds^2_8.
\end{align}

Following our definition \cite{Afrasiar:2024ldn}, the timelike entanglement entropy (tEE) reads as 
\begin{align}
\mathcal{S}^{(tEE)}&=\frac{1}{G_N}\int d^{9}x e^{- 2 \phi}\sqrt{\det g_9}\nonumber\\
&=\frac{\pi \rho^3_0}{6G_N}V_{CY_2}V_\psi \int_0^\infty dr \sqrt{1-\cosh^2 r t'^2(r)}\nonumber\\
&=\frac{\pi \rho^3_0}{6G_N}V_{CY_2}V_\psi \int_0^1\frac{dz}{z}\sqrt{f(z)-t'^2(z)}
\label{e3.10}
\end{align}
where we replace $ \cosh r=\frac{1}{z} $ and the function $ f(z) $ is defined in \eqref{e2.8}.

\begin{figure}
\begin{center}
\includegraphics[scale=0.5]{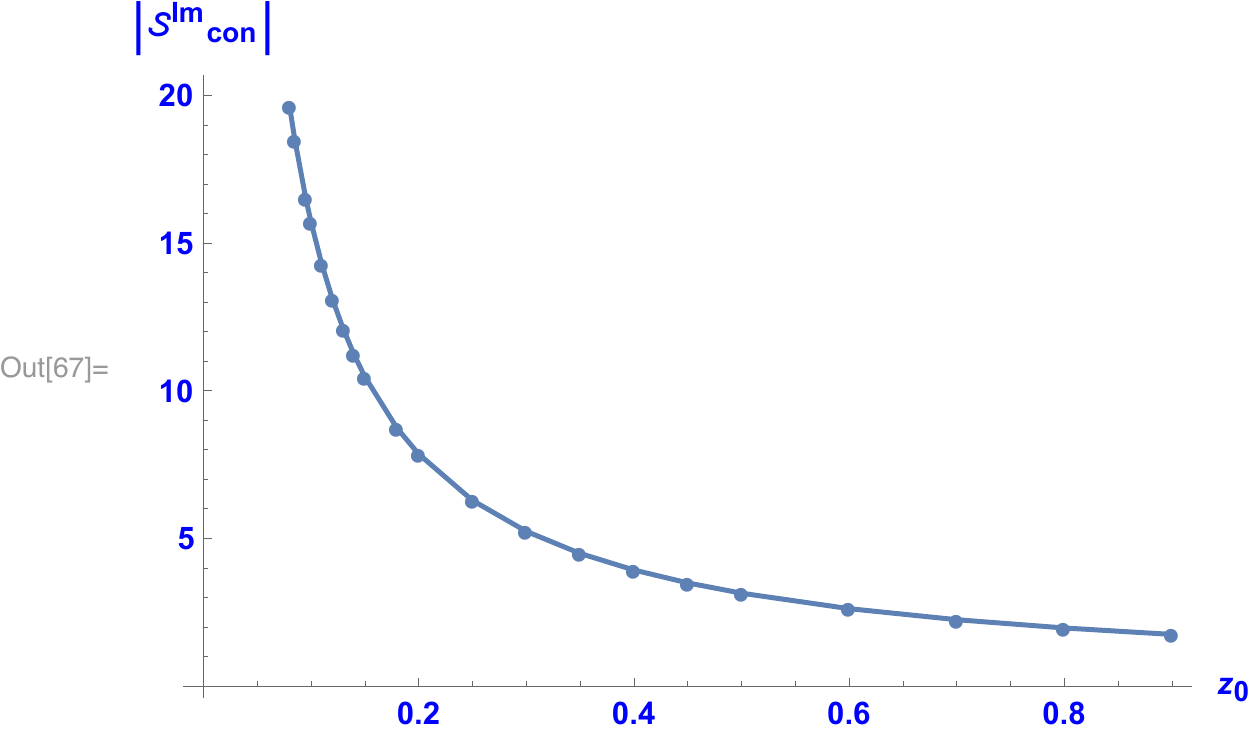}
  \caption{(a) RG flow of tEE \eqref{e3.13} in $ \mathcal{N}=4 $ SCQM.} \label{scqm}
  \end{center}
\end{figure}

The equation of motion for $ t(z) $ turns out to be
\begin{align}
\label{e3.11}
t'^2(z)=\frac{C^2 z^2 f (z)}{1+C^2 z^2}
\end{align}
where $ C $ is the constant of integration.

Substituting \eqref{e3.11} into \eqref{e3.10}, we finally obtain
\begin{align}
\mathcal{S}^{(tEE)}=\frac{\pi \rho^3_0}{6G_N}V_{CY_2}V_\psi \int_0^1\frac{dz}{z}\frac{\sqrt{f(z)}}{\sqrt{1+ C^2 z^2}}
\end{align}
where $ \rho_0 $ defines the size of the $ \mathcal{N}=4 $ quiver. For NATD example, the superconformal quantum mechanical quiver is unbounded. This is typically reflected in the dual gravity description by setting the upper limit of the holographic ($ \rho $) axis to be large enough namely, $ \rho_0 \gg 1 $. Similar remarks hold for the matrix model calculations in the previous section.

Like in the previous example, turning point (at $ z=z_0 $) for the connected extremal surface in the bulk corresponds to setting $ C^2=-\frac{1}{z^2_0} $, which finally yields $\mathcal{S}^{(tEE)}=z_0 \mathcal{S}^{(Im)}_{con}$ where
\begin{align}
\label{e3.13}
\mathcal{S}^{(Im)}_{con}&=\frac{-i\pi \rho^3_0}{3G_N}V_{CY_2}V_\psi \int_{z_0}^1\frac{dz}{z}\frac{\sqrt{f(z)}}{\sqrt{ |z^2 - z^2_0|}}.
\end{align}

As a side remark, the subsystem size turns out to be
\begin{align}
t_{sub}=2 \int_{z_0}^1 dz \frac{z\sqrt{f(z)}}{\sqrt{ |z^2 - z^2_0|}}.
\end{align}

The integral above in \eqref{e3.13} dictates the RG flow of tEE, as in the previous example. For eaxmple, taking the UV limit of the turning point $ z_0 \sim \epsilon $ in the above integral \eqref{e3.13}, one finds\footnote{Here, $z_0 $ acts like a UV cut-off for the dual superconformal field theory. In other words, we have a CFT with a UV cut-off. When we describe a bulk RG flow, this essentially corresponds to the fact that one is integrating out all the degrees of freedom above this cut-off, which therefore leads to an \emph{effective} description of a CFT.}
\begin{align}
\label{e3.15}
|\mathcal{S}^{(Im)}_{con}|_{UV}=\frac{\pi \rho_0^3}{3G_N}V_{CY_2}V_\psi \int_{\epsilon}^1\frac{dz}{z^2}\frac{1}{\sqrt{1- z^2}}=\frac{\pi \rho_0^3}{3G_N}V_{CY_2}V_\psi \frac{1}{\epsilon}
\end{align}
which exhibits a pole of order one, near the UV asymptotic ($ z \sim \epsilon $). This is also depicted in Fig.\ref{scqm}, which clearly indicates the RG flow of tEE \eqref{e3.13} starting from UV asymptotic ($ z \sim \epsilon $) to deep IR ($ z \sim 1 $), which is identical to what has been observed for matrix models in ($ 0+1 $)d. 
\subsection{Mapping into the central charge}
Calculation of the central charge \cite{Macpherson:2014eza}, \cite{Bea:2015fja}-\cite{Chatzis:2024kdu}  follows steps identical to those preformed in the case of matrix models. To begin with, we express 10d metric \eqref{e3.1} as
\begin{align}
&ds^2_{10}=\frac{1}{4}(- \cosh^2r dt^2 +dr^2)+\mathcal{G}_{ij}d\omega^i d\omega^j\nonumber\\
&\mathcal{G}_{ij}d\omega^i d\omega^j\equiv ds^2_8= f_2(\rho)(d \theta^2 +\sin^2\theta d\xi^2)+f_3(\rho)ds^2_{CY_2}+f_4(\rho)(d\rho^2 + d\psi^2).
\end{align}

Like in the case of matrix models, the contribution to the holographic central charge ($ c_{hol} $) for ($ 0+1 $)d QFTs comes through the volume measure of the internal eight dimensional manifold  
\begin{align}
\label{e3.17}
c_{hol}&= \frac{1}{G_N}\int d^8 \omega e^{-2\phi}\sqrt{\det \mathcal{G}_{ij}}\nonumber\\
&=\frac{\pi}{G_N}V_{CY_2}V_\psi \int_0^{\rho_0}d\rho \rho^2
\end{align}
which by virtue of \eqref{e3.15}, yields the following relation
\begin{align}
|\mathcal{S}^{(Im)}_{con}|_{UV}=\frac{c_{hol}}{\epsilon}.
\end{align}
In other words, tEE \eqref{e3.13} in the limit in which the turning point ($z_0$) reaches the UV cut-off $ z \sim \epsilon $, matches with the central charge \eqref{e3.17} of the dual $ \mathcal{N}=4 $ SCQM living at the boundary.
\subsection{Comments on complexity}
Calculations on complexity follows identically as in the case of matrix models. To begin with, we express the 10d metric as follows
\begin{align}
ds^2_{10}=g_{tt}dt^2 +ds^2_9
\end{align}
where the nine dimensional subspace has a metric of the form
\begin{align}
ds^2_9=\frac{1}{4}dr^2 +ds^2_8.
\end{align}

Following the CV proposal  \cite{Stanford:2014jda}, the holographic complexity \cite{Chatzis:2024kdu}, \cite{Fatemiabhari:2024aua} reads as
\begin{align}
\mathcal{C}_V&=\frac{1}{G_N}\int d^9x e^{-2 \phi}\sqrt{\det g_9}\nonumber\\
&=\frac{c_{hol}}{2}\int_{0}^{\infty} dr\nonumber\\
&=\frac{\epsilon}{2}|\mathcal{S}^{(Im)}_{con}|_{UV}\int_\epsilon^1 \frac{dz}{z}\frac{1}{\sqrt{1-z^2}}\nonumber\\
&=\frac{\epsilon}{2}\log \Big(\frac{2}{\epsilon}\Big) |\mathcal{S}^{(Im)}_{con}|_{UV}.
\end{align}
\section{Conclusions and Outlook}
We now summarise key findings of our paper. Both imaginary and real components of tEE diverge in the UV asymptotic, which is an artefact of large area contribution to the tEE.  By introducing a pair of disconnected surfaces, one could tame these UV divergences in tEE, which we denote as the area renormalisation of tEE. Interestingly enough, the imaginary component of tEE \eqref{e2.15} exhibits a property that is quite similar in spirit as that of a holographic $ c $ function, namely it vanishes in the deep IR and diverges in the asymptotic infinity.

We probe further into it by explicitly constructing the holographic flow central charge following the lines of \cite{Macpherson:2014eza}, \cite{Bea:2015fja}-\cite{Chatzis:2024kdu}, which shows a remarkable similarity in the pole (of order $3$) structure of both tEE \eqref{e2.15} and the $c$ function \eqref{e2.33} near the UV asymptotic of the 10d spacetime. This clearly indicates that the \emph{imaginary} component of tEE \eqref{e2.15} indeed counts the number of degrees of freedom in ($ 0+1 $)d matrix model during its RG flow from UV to IR. We further confirm this claim, where we calculate complexity in the matrix model which (is a measure of number of degrees of freedom) exhibits a pole identical to that of tEE \eqref{e2.15} near the UV asymptotic. Finally, we estimate (regularised) tEE \eqref{e2.46} in Euclidean signature which exhibits identical characteristic as that of \eqref{e2.15} in the Lorentzian signature.

We repeat these calculations for a new class of $ \mathcal{N}=4 $ SUSY conformal quantum mechanical model (SCQM) in ($ 0+1 $)d, that are dual to a class of type IIB backgrounds with an $ AdS_2 $ factor \cite{Lozano:2020txg}. An estimation of tEE \eqref{e3.13} for these superconformal quantum mechanical quivers reveals identical features as found in the matrix model example namely, tEE decreases in a RG flow from UV to deep IR which is further identified with the holographic central charge ($ c_{hol} $) as well as mapped into the complexity associated with the dual SCQM quiver. 

To summarise, tEE (in Lorentzian signature) is a good measure of the number of degrees of freedom during RG flow from UV to deep IR, in a class of SQFTs that exist in ($ 0+1 $)d. As a natural consequnce of this, tEE can be mapped into other QFT observables namely the flow central charge and the complexity, which also counts the degrees of freedom in a RG flow. 

This paper outlines two calculations which are indeed model dependent. It would be really nice to extend the above ideas and prove the connection between the $ c $ function and the tEE for more generic QFTs living in ($ 0+1 $)d and in particular in a model independent way. It would be really nice to explore similar connection between different field theory observables for QFTs living in higher dimensions. We hope to address some of these issues in near future.
\subsection*{Acknowledgments}
It is a pleasure to thank Carlos Nunez for several useful discussions. The author is indebted to the authorities of IIT Roorkee for their unconditional support towards researches in basic sciences. The author also acknowledges The Royal Society, UK for financial assistance. The author also acknowledges the Mathematical Research Impact Centric Support (MATRICS) grant (MTR/2023/000005) received from ANRF, India.


\end{document}